# On observation of dispersion in tunable second-order nonlinearities of silicon-rich nitride thin films

Hung-Hsi Lin,[1*], Rajat Sharma,[2*†] Alex Friedman,[2†] Benjamin M. Cromey,[3]
Felipe Vallini,[2] Matthew W. Puckett,[2] Khanh Kieu,[3] and Yeshaiahu Fainman[2]

[1]*Materials Science and Engineering, University of California, San Diego, 9500 Gilman Drive, La Jolla, California 92093, USA*
[2]*Department of Electrical & Computer Engineering, University of California, San Diego, 9500 Gilman Drive, La Jolla, CA 92093, USA*
[3]*College of Optical Sciences, The University of Arizona, 1630 E University Blvd, Tucson, AZ 85719, USA*

**Abstract**

We present experimental results on second-harmonic generation in non-stoichiometric, silicon-rich nitride (SRN) films. The as-deposited film presents a second-order nonlinear coefficient, or $\chi^{(2)}$, as high as 8pm/V. This value can be widely tuned using the electric field induced second harmonic effect (EFISH), and a maximum value of 22.7pm/V was achieved with this technique. We further illustrate that the second-order nonlinear coefficient exhibited by these films can be highly dispersive in nature, and requires further study and analysis to evaluate their viability for in-waveguide applications at telecommunication wavelengths.

**Introduction**

A highly desirable feature from an integrated photonics platform is to provide an efficient on-chip solution to wave-mixing and modulation applications. Unfortunately, the most commercially successful material for integrated photonics, silicon, being centrosymmetric, does not exhibit a non-zero second-order nonlinearity and hence cannot address these applications [1, 2]. As such, several alternative approaches have been proposed and demonstrated to achieve these functionalities in the silicon photonics platform. These include approaches based on strain-induced nonlinearities, the free-carrier plasma dispersion effect and heterogenous integration of materials like lithium niobite [3-5]. However, these approaches have their own shortcomings, including but not limited to, the inherent nonlinear nature of the achieved modulation, and a lack of CMOS compatibility. Because of these limitations, there is an acute need for a highly nonlinear material that can be integrated with silicon to achieve efficient, high-speed, and linear electro-optic modulation. In recent years, silicon nitride has emerged as one such promising candidate with salient features such as a wide transparency window, ease of fabrication and compatibility with silicon photonics manufacturing.

There have been reports in the literature of thin-films of silicon nitride exhibiting a bulk second-order nonlinearity with a value of the nonlinear coefficient, $\chi^{(2)}$, as high as ~3pm/V, at a pump wavelength of 800nm [6, 7]. In addition, in-waveguide measurements have also been reported, demonstrating this phenomenon through phase-matched second-harmonic generation, where interestingly the value of the reported nonlinear coefficient was as low as ~0.3pm/V at a pump wavelength of 1550nm [7]. The authors have attributed this discrepancy to reasons such as a lack of a perfect phase-matching and/or modifications introduced into the materials during the fabrication process. However, to the best of our knowledge, there has not been any discussion of or reported results on this discrepancy in the literature.

---

[*] *The authors contribute equivalently*
[†] *Corresponding author: r8sharma@eng.ucsd.edu, amfriedm@eng.ucsd.edu*

In this manuscript, we undertake a systematic evaluation of the second-order nonlinearity exhibited by silicon nitride films. We discuss methods to enhance the observed nonlinearity by increasing the silicon content in the films as well as through the electric-field induced second-harmonic effect (EFISH). Lastly, we report on our observation of a high degree of dispersion in the second-order nonlinearity exhibited by these films as a function of wavelength.

## Silicon-rich nitride

Bulk nonlinearities in silicon nitride thin films deposited through plasma enhanced chemical vapor deposition, PECVD, and RF magnetron sputtering have been reported on in the past [6]. The first measurement of this nonlinearity using in-waveguide experiments was reported in 2016 [7], where the authors carried out phase-matched second-harmonic generation in stoichiometric silicon nitride waveguides. The magnitude of the reported nonlinearity was low with in-waveguide measurements (at 1550nm) yielding values of $\chi^{(2)}$ lower than 1pm/V. These values have subsequently been reproduced and confirmed in a separate measurement by Billat et al [8]. By leveraging a unique attribute of the silicon nitride platform, this relatively small nonlinearity can be enhanced by changing the stoichiometry of the deposited films. It is known that increasing the silicon content in sputtered silicon nitride thin films, yielding so-called silicon-rich nitride (SRN) films, leads to an enhancement in the magnitude of the third-order nonlinearity [9]. However, these films suffer from high propagation loss, making them inapplicable to many in-waveguide applications. There have been two recent reports [10,11] extending these results to PECVD deposited SRN films carried out using free-space measurements with pump wavelengths of 800nm and 1040nm respectively. The SRN material was shown to possess an enhanced second-order nonlinearity compared to that of stoichiometric films, while its propagation loss values remained relatively low.

With an intent to evaluate these films for in-waveguide applications, we carried out a systematic study of the effect of silicon content on the exhibited linear and nonlinear optical properties of the films. Three different samples, labelled $S_1$, $S_2$, and $S_3$, were fabricated with silicon nitride films deposited on fused-silica substrates. The flow-rate of silane ($SiH_4$), one of the precursors in the PECVD process, was varied across the samples from 180, 276, and 500sccm, while keeping all the other deposition parameters (outlined in [7]) constant. Ellipsometry measurements using light at 632nm confirmed that this fabrication processes produced films with unequal refractive indices of 1.9 ($S_1$), 2.08 ($S_2$) and 2.25 ($S_3$), with the index of the film scaling positively with the $SiH_4$ flow-rate. The reason for this was confirmed to be an increase in silicon content by carrying out electron-dispersive X-ray (EDX) spectroscopic measurements. Fig. 1a shows the composition of the films in terms of their silicon and nitride atomic percentages (shown in red and blue colors respectively), with a silicon content of 41% ($S_1$), 51% ($S_2$), and 56% ($S_3$) across the three films.

## Free-space measurements

In order to characterize the effect of the silicon content in these films on their nonlinear properties, polarization dependent SHG experiments were carried out using a femtosecond Ti:Sapphire laser source at a wavelength of 800nm, with a pulse duration of 150fs, an 80MHz repetition rate, and

100mW of average power. Second-harmonic signals, generated for both s- and p-polarized pumps at a wavelength of 800nm, were measured as a function of the polarization angle of the pump, and are shown in Fig. 1b. The revised Maker fringes analysis was then employed to carry out the tensorial analysis of the second-order nonlinearity [7, 11-13]. As is evident from the figure, the generated second-harmonic signal from the film with the highest silicon content ($S_3$) is up to 20 times larger compared to that generated from the film with a stoichiometric composition ($S_1$). Additionally, these polarization dependent second-harmonic responses were then used to extract the tensorial components of the observed $\chi^{(2)}$ in these films. The extracted values of the all-normal ($\chi^{(2)}_{zzz}$) and in-plane ($\chi^{(2)}_{zxx}$) components are tabulated in Table I, and are found to be up to 3.3 times larger for the SRN film (sample $S_3$) when compared with the stoichiometric film (sample $S_1$). Furthermore, to the best of our knowledge the measured tensor component $\chi^{(2)}_{zzz}$ in the SRN film is the largest reported to-date in as-deposited PECVD silicon nitride films.

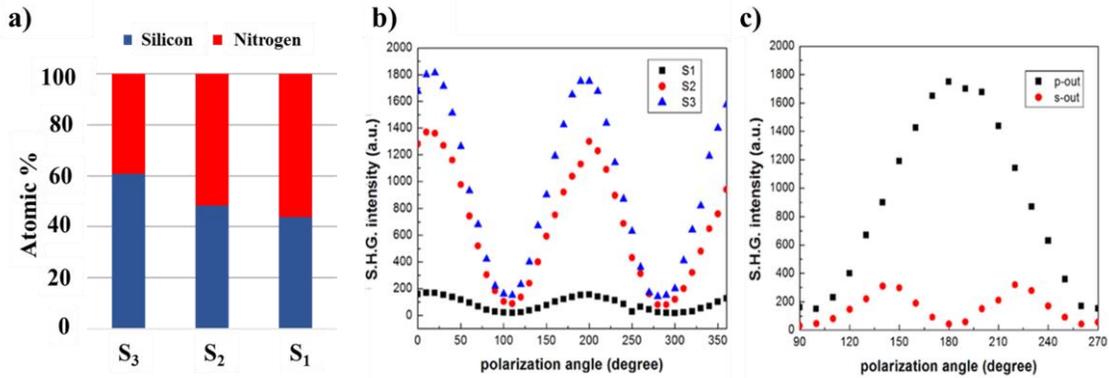

FIG. 1. (a) Composition of the samples, $S_1$, $S_2$, and $S_3$, in terms of the atomic percentages of silicon and nitrogen, extracted using EDX spectroscopy (b) Second-harmonic signal generated from the three films as a function of polarization angle for p- polarization. (c) SHG signals from the SRN film ($S_3$) for both s- and p-polarized input pumps.

TABLE I. The depostion recipes and measured properties of silicon nitride thin films.

| Sample | SiH$_4$ flow rate (sccm) | Si % | Refractive index | $\chi^{(2)}_{zzz}$ (pm/V) | $\chi^{(2)}_{zxx}$ (pm/V) |
|---|---|---|---|---|---|
| $S_1$ | 180 | 41 | 1.9 | 2.4 | 0.4 |
| $S_2$ | 276 | 51 | 2.08 | 5.8 | 1.1 |
| $S_3$ | 500 | 56 | 2.25 | 8 | 1.9 |

### Electric field induced enhancement in nonlinearities

An alternative method to enhance the nonlinear response in these films is the electric-field induced second-harmonic (EFISH) effect. In our previous work [7], we presented in-waveguide results demonstrating enhancement of the second-harmonic response from silicon nitride waveguides by applying an electric field across them. It was shown then that the applied external electric field interacts with the third-order nonlinearity of the films, resulting in a higher effective second-order response, and a corresponding increase in the intensity of the second-harmonic signal from the waveguides [7]. To analyze the tunability of the nonlinear response in the SRN films, we fabricated 50nm thick films of the three silicon nitride films sandwiched between two electrodes made of

10nm thick layers of indium tin oxide (ITO) on a fused silicon substrate as shown in the schematic in Fig. 2a. The ITO films serve as the transparent electrodes across which an external DC voltage is applied while carrying out SHG experiments. It should be noted that while the ITO films were not optimized with respect to their conductance, they were still sufficient for carrying out preliminary studies on the EFISH induced tunability in the films. Fig. 2b shows the SHG response for the SRN film as a function of the applied voltage, clearly demonstrating a wide range of tunability. The SHG response is found to increase quadratically with respect to the applied voltage, which is in agreement with our prediction of EFISH. It should also be noted that the minimum in the measured quadratic response is found to be at a small negative bias voltage and not at zero-bias. This is because at this voltage, the artificially created second-order nonlinearity generated by the external electric field and the high $\chi^{(3)}$ coefficient of SiN perfectly negates the all-normal $\chi^{(2)}_{zzz}$ present in the as-deposited film. As a result, the magnitude of the SHG response reduces, but is still not perfectly zero because of contributions from the in-plane components of the $\chi^{(2)}$ in the as-deposited nitride film, ITO layers, as well as those arising out of any surface nonlinearities.

The SHG signal measured at an applied voltage of 8V was found to be up to 36 times larger than that measured at zero-bias. Table II summarizes the calculated highest and lowest effective $\chi^{(2)}_{zzz}$ components of three different silicon nitride samples fabricated with the same silicon contents as $S_1$, $S_2$ and $S_3$. The SRN film demonstrates an effective $\chi^{(2)}$ spanning from ~3pm/V to as high as 22pm/V. This relatively large range of tunability is in accordance with reports in the literature of an enhanced third-order nonlinear response in silicon-rich nitride films when compared to their stoichiometric counterparts [14,15]

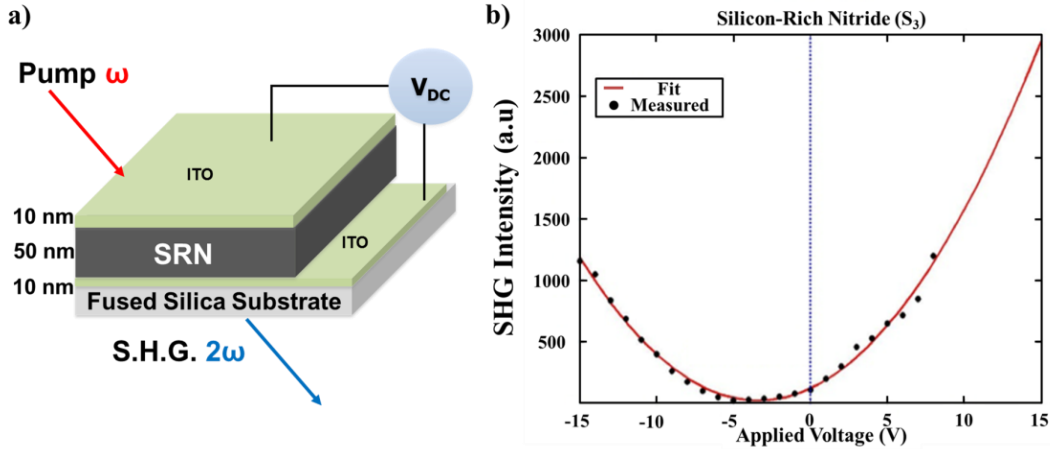

FIG. 2. (a) Schematic illustration of the SHG response from a 50nm silicon nitride film sandwiched between two 10nm ITO films on a fused silica substrate. (b) Experimental (black dots) and quadratic fitting (red) curve showing the SHG response from a SRN sample as a function of the external voltage applied across the layer.

TABLE II. As-deposited, lowest and highest effective $\chi^{(2)}$ of three different silicon nitride films with different silicon contents. The highest range of tunability is exhibited by the SRN film with a value of $\chi^{(2)}$ as high as 22.7pm/V.

| Sample | SiH$_4$ flow rate (sccm) | $\chi^{(2)}_{eff}$ (no bias) (pm/V) | Lowest $\chi^{(2)}_{eff}$ (pm/V) | Highest $\chi^{(2)}_{eff}$ (pm/V) |
|---|---|---|---|---|
| $S_1^*$ | 180 | 2.6 | 2.1 | 3.9 |
| $S_2^*$ | 276 | 3.35 | 2.39 | 5.93 |
| $S_3^*$ | 500 | 7.5 | 3.9 | 22.7 |

# Dispersion in the observed nonlinearity

The bulk of the studies on second-order nonlinearities in silicon nitride films have been carried out using either 800nm or 1064nm sources [6, 11]. While the values of nonlinearities reported using these pump wavelengths are relatively high, other works which pursue in-waveguide experiments at a pump wavelength of 1550nm have measured lower values of the nonlinearity [7, 8]. The cause of this discrepancy between values measured using two different pump-wavelengths was attributed previously to a lack of perfect phase-matching in waveguide SHG experiments carried out using a pump at 1550nm.

To test for wavelength dependence in the nonlinearities of our films, we carried out reflection SHG measurements on all three samples, using pumps at 1040nm and 1550nm. These two wavelengths were chosen solely based on source availability. The exact methodology employed in carrying out these measurements is detailed in the supplementary material, while the optical setup is shown in Fig. 3a. In order to make a fair comparison, parameters such as incident angle, spot size of the beam, average pump power, and polarization state were kept constant while carrying out the measurements across the three samples. Fig. 3b shows the magnitude of the SHG signal, corresponding to the polarization with the maximum second-harmonic signal, generated from the three samples at the two pump-wavelengths. Sample $S_3$, with the highest silicon content, demonstrates the largest second-harmonic signal among the three samples at both pump wavelengths. Interestingly, there is a significant reduction in the SHG intensity from 1040nm to 1550nm of up to 20-30 times across all three samples, implying a reduction of ~5 times in the effective second-order nonlinear coefficient.

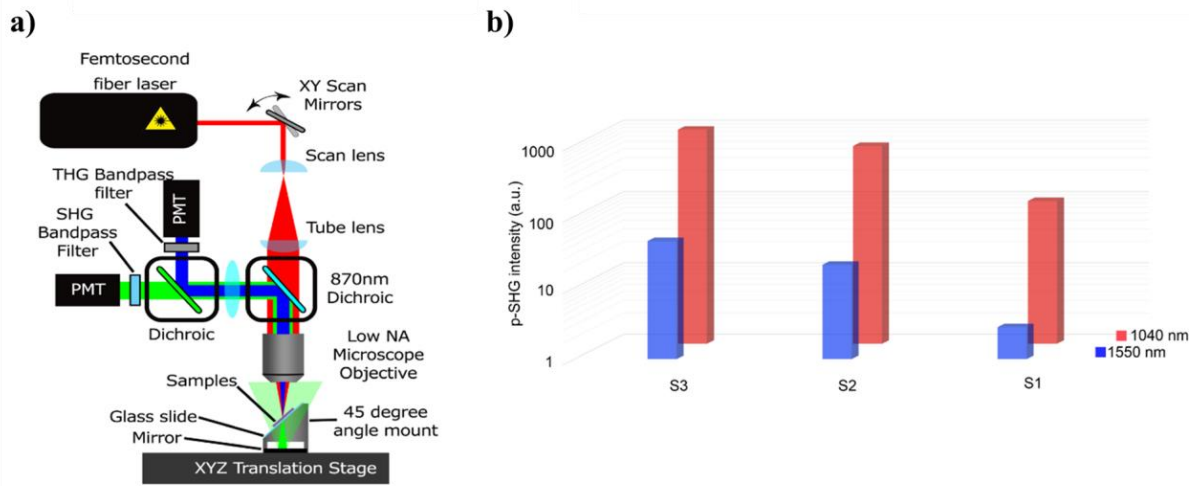

FIG. 3. (a) Schematic of the reflective second-harmonic generation setup. (b) The generated reflective p-polarized SHG signals from the three samples, corresponding to pump wavelengths of 1040 nm (red bars) and 1550 nm (blue bars).

Dispersion in nonlinear susceptibilities is a well-known phenomenon explained by Miller's rule, which defines a relation between dispersion in refractive index and dispersion in second-order nonlinearities [17,18]. Furthermore, work in literature has shown a correlation between increasing silicon content in silicon nitride films and the dispersion in refractive index of such films [19]. These two facts together imply that dispersion in $\chi^{(2)}$ is expected from 1064nm to 1550nm and expected to be the highest in the film with the highest silicon content. However, the magnitude of dispersion demonstrated in this work far exceeds that predicted by Miller's rule and therefore the exact origins of this remains to be explained. Specifically, a difference in magnitude of second harmonic generation between 1064nm and 1550nm of 20 to 30 times is seen across all three films, implying a reduction in $\chi^{(2)}$ of up to 5 times.

## In-waveguide loss characterizations

To characterize the viability of SRN films for on-chip applications, in-waveguide loss measurements were carried out on SRN waveguides on oxide-on-silicon substrates. The thicknesses of the SRN device layer, deposited using PECVD, and that of the oxide below were 430nm and 3μm respectively. Ring-resonators coupled to bus waveguides were then fabricated using a combination of electron beam lithography and inductively coupled plasma reactive ion etching (ICP-RIE) using a $C_4F_8$, $SF_6$ plasma as outlined in [7]. The waveguide widths were kept at 1000nm, and the coupling gaps were varied from 150-300nm, in the case of $TE_0$, and from 500-700nm, in the case of $TM_0$ transmission to achieve critical coupling. The choice of the width and height was made keeping in mind the need to have single-mode operation of both TE- and TM polarized modes at a wavelength of 1550nm. In-waveguide measurements were then performed using a fiber-in, free-space out setup with an Agilent 8164-B tunable CW laser source spanning a wavelength range of 1.46 to 1.64μm [7,16]. Fig. 4 shows the measured normalized transmission (dB) of the ring-resonators for TE and TM polarizations for the cases closest to the critical coupling regime. This was achieved in the case of TE transmission at a coupling gap of 250nm, and at a gap of 600nm for TM transmission. Also shown are the calculated modal propagation loss values, α (dB/cm), extracted from each resonance individually via fitting to a well-known Lorentzian function for transmission of a bus coupled ring-resonator as outlined in [16]. The propagation loss values for the TE and TM cases, calculated from resonances at 1524.7nm and 1563.9nm, were found to be 9.54 and 14.60 dB/cm respectively. These values, while relatively high when compared to stoichiometric silicon nitride waveguides, are close to other in-waveguide loss measurements for SRN films in literature [14]. Furthermore, these can be improved by optimizing the fabrication process such as the etching recipe and/or employing post-deposition annealing, as well as by increasing the width of the waveguide to reduce sidewall scattering [7].

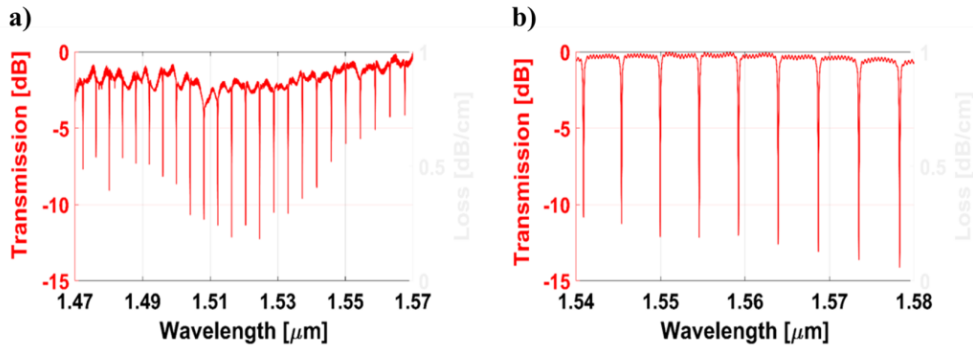

FIG. 4. (a) Red - Normalized transmission, across a wavelength range of 1.47 to 1.57μm, of the $TE_0$ mode in a 1um wide SRN waveguide coupled to a ring-resonator at a coupling gap of 250nm. Blue – Loss extracted from each resonance by fitting to a Lorentzian line-shape function for transmission of a ring-resonator [16]. (b) Red - Normalized transmission, across a wavelength range of 1.54 to 1.58μm, of the $TM_0$ mode in a 1um wide SRN waveguide coupled to a ring-resonator at a coupling gap of 600nm. Blue – Loss extracted from each resonance.

# Conclusions and Discussion

Historically, bulk second-order nonlinearities in silicon nitride films have been an anomalous finding, and as such, their exact origins have been ambiguous. Furthermore, the variety of techniques employed to deposit and characterize these films leads to further confusion in the reported properties of such films. It is important to make a fair comparison of these films in terms of deposition technique, wavelength of characterization, and silicon content.

In our past work we have reported a large discrepancy of up to an order of magnitude in the nonlinear coefficient of stoichiometric nitride films. These were extracted in free-space at 800nm and in-waveguide at 1550nm, using phase-matched SHG. In this case, we hypothesized that the discrepancy was due to a lack of perfect phase-matching and/or material changes as a result of processing during waveguide fabrication [7]. Similarly, recent independent results by Billat et al and Porcel et al [8,20] demonstrated all-optical quasi-phase matched SHG in stoichiometric silicon nitride waveguides, reporting a value of 0.3pm/V (at 1550nm) and 3.7pm/V (at 1064nm) respectively. These results achieved quasi-phase matching due to an all-optical, charge separation-induced grating in the silicon nitride waveguides, via the coherent photogalvanic effect (CPE). The latter study attributes the higher coefficient observed to the nature of the nitride film deposited and compares these results to other reported coefficients including those in Ref. 7. However, care was not taken to account for the wavelength employed in each of the studies to which the authors compare their result. Considering the results in our current study, we attribute the large discrepancy in these references to the dispersion in the nonlinearity of stoichiometric nitride thin films as opposed to an inherently larger coefficient. This observed dispersion, it should be noted, is much higher than what can be attributed to Miller's rule [17], which relates the relative magnitude of the nonlinear susceptibilities to the linear susceptibilities at the respective wavelengths. The exact origin of this high dispersion remains. Finally, this study's use of thin-films was carefully chosen to be performed in free space to avoid the need for phase matching of any kind, neither all optical quasi-phase matching nor modal dispersion-based phase matching.

In summary, this work has clearly demonstrated that increasing silicon content in PECVD deposited silicon nitride films results in enhancement in the second-order nonlinearity of such films, achieving a value as high as 8pm/V in as-deposited SRN films. We then demonstrate a relatively large tunability range of the nonlinear coefficient, with a highest demonstrable coefficient of 22pm/V, using the EFISH effect. Furthermore, we demonstrate that the inherent nonlinearity in as-deposited films is highly dispersive, not just in the case of silicon-rich compositions, but also in the case of stoichiometric films. Finally, it is our opinion that the highly dispersive nature of the second-order nonlinearity in these films should be taken into consideration by any future studies comparing the nonlinearity and/or evaluating the viability of such films for on-chip nonlinear and or electro-optic applications.

# Acknowledgements

This work was supported by the Defense Advanced Research Projects Agency (DARPA), the National Science Foundation (NSF), the NSF ERC CIAN, NSF's NNCI San Diego Nanotechnology Infrastructure (SDNI), the NSF Graduate Research Fellowship under DGE-1143953, the Office of Naval Research (ONR) Multidisciplinary University Research Initiative (MURI), the Army Research Office (ARO), and the Cymer Corporation. We thank UCSD's staff Ryan Anderson for the discussion on the equipment for electrical characterization.

# Supplementary materials

### Dispersion study

A multiphoton microscope setup was used to examine the samples at 1040 nm and at 1550 nm [1]. A home-built angled mount was placed at the focus of the microscope in order to hold the sample at an angle of 45 degrees relative to the incoming beam. A mirror was placed beneath the mount to maximize the amount of signal that was epi-detected by reflecting any forward phase matched SHG light back into the objective. The SHG signals were detected by a pair of Hammamatsu PMTs (10721-20 for SHG). The outputs from these PMTs were then amplified by preamplifiers (Stanford Research SR570) and detected via a National Instruments DAQ board and lab-developed LabVIEW software.

The lasers used were lab-built femtosecond fiber lasers based on a single walled carbon nanotube saturable absorber [2-3]. The unpolarized output of the 1040 nm laser was transformed into linear polarization with a polarizer, and a half wave plate (HWP) was used to rotate the polarization plane. A 410 nm dichroic mirror separated the SHG light at 520 nm to the PMT. A HWP was used to rotate the polarized output of the 1550 nm laser to achieve the desired polarization plane. A dichroic at 560 nm and filters at 517 nm and at 780 nm were used to isolate the SHG and THG signals.

For both lasers, the quartz sample was imaged first in order to provide a reference of the signal intensity for SHG. First, an image was captured using lab developed software. The voltage detected by the DAQ was recorded for each pixel, which could be converted back to current by the gain setting of the preamplifier. Additionally, the output of the PMT's were sent directly to a picoammeter (Keithley 6485), and the current values recorded. This could be converted to power using the spectral response curves from the Hamamatsu PMTs. The noise floor was recorded for each measurement. Measurements were taken both for s and p polarization for all samples and both lasers.